\documentclass[12pt]{iopart}
\usepackage{graphicx}
\usepackage{epsfig}
\usepackage{iopams}

\begin{document}

\title[Generating multipartite entanglement]{Generating genuine multipartite entanglement via XY-interaction and via projective measurements}

\author{Mazhar Ali$^{1,2}$}
\address{$^1$Naturwissenschaftlich-Technische Fakult\"at, Universit\"{a}t Siegen, Walter-Flex-Stra\ss e 3, 57068 Siegen, Germany \\
$^2$Department of Electrical Engineering, COMSATS Institute of Information Technology, 22060 Abbottabad, Pakistan} 
\ead{mazharaliawan@yahoo.com}

\begin{abstract}
We have studied the generation of multipartite entangled states for the superconducting phase qubits. The experiments performed 
in this direction have the capacity to generate several specific multipartite entangled 
states for three and four qubits. Our studies are also important as we have used a 
computable measure of genuine multipartite entanglement whereas all previous studies analyzed certain probability amplitudes.  
As a comparison, we have reviewed the generation of multipartite entangled states via von Neumann projective measurements. 
\end{abstract}


\maketitle

\section{Introduction}\label{S:Int}

Quantum entanglement has been recognized as a resource with applications in the emerging field of quantum information and quantum computation 
\cite{NC-QIQC-2000, Horodecki-RMP-2009}. The creation and measurement of entangled states is crucial for the various physical implementations of 
quantum computers \cite{Ladd-Nature464-2010, Buluta-RPP74-2011}. Therefore it is of much interest and importance to generate entanglement in 
experiments and also characterize it in theory. The description of quantum entanglement for bipartite quantum systems is relatively simple as any 
given quantum state is either entangled or separable. However for quantum systems with more than two subsystems, this problem becomes richer and also 
more difficult. For the simplest multipartite quantum system of three qubit system, it was shown that three qubits can be entangled in two 
fundamentally different ways under stochastic local operations and classical communication (SLOCC) \cite{Duer-PRA62-2000}. In contrast, such 
inequivalent SLOCC-classes of entangled states for four qubits are already infinite \cite{Verstraete-PRA65-2002, Lamata-PRA75-2007}.     

One of the basic task in a quantum computer is the implementation of a set of universal gates usually a two-qubit gate such as controlled-NOT (CNOT) 
gate and single qubit rotations \cite{Barenco-PRA52-1995}. However, it is also possible that a three-qubit gate like Toffoli gate achieves 
universality \cite{Shi-QIC3-2003, Lanyon-NP5-2008, Monz-PRL102-2009}. Therefore it is desired to design experiments with a direct 
implementation of multi-qubit gates. Recently, such implementations have been successfully performed on the superconducting phase qubits 
\cite{Gali-PRA78-2008, Neeley-Nature467-2010, Matsuo-JPSJ-2007, Tsomokos-NJP10-2008}. In these experiments, the phase qubits were 
coupled by connecting them with a capacitor to generate multi-qubit interactions leading to multi-qubit gates rather than designing them from 
more elementary two-qubit gates. Moreover, it was shown that one can also use two qubit gates to create $|GHZ\rangle$ states and a more efficient 
entangling protocol based on a single three qubit gate to create $|W\rangle$ state \cite{Neeley-Nature467-2010}. We have extended these studies 
for three and four qubits and have shown that this experimental setup can offer some unique possibilities which are not explored till now. We have 
also proposed an experimental architecture for generating four qubit $|\chi_4\rangle$ state which might be implemented with small modification in 
the original experiment.

Another approach to create SLOCC-inequivalent multipartite entangled states is to apply von Neumann projective measurements on some of the 
subsystems \cite{Ionicioiu-PRA78-2008, Kiesel-PRL98-2007, Wieczorek-PRA79-2009, Wieczorek-PRL103-2009}. This way to create entangled states is due to 
the property of certain symmetric multipartite entangled states that allow a more flexible preparation of families of SLOCC-inequivalent entangled 
states by projective measurements on small subsystems. We have reviewed this method to compare the results. We have shown that it is possible to 
create both $GHZ$-type and $W$-type states of three qubits by applying projective measurements on a single qubit of four qubit states. We have shown 
that in creating $|\chi_4\rangle$ state, we have used $GHZ$-type entanglement and $W$-type entanglement. Although these three states are 
inequivalent, nevertheless, in the reverse process of projective measurements, we can extract $GHZ$-type and $W$-type entanglement from 
$|\chi_4\rangle$ state. This process of creating inequivalent entangled states has increased our understanding about multipartite entangled states.

This paper is organized as follows. In Section \ref{S:GME}, we review the concept of genuine multipartite entanglement and describe the computable 
measure which detects such type of entanglement. In Section \ref{S:GXY}, we study the possibility to generate various multipartite entangled states. 
We reexamine the possibility to generate genuine multipartite entanglement via von Neumann measurements in Section \ref{S:GVN}. Finally we offer 
some conclusions in Section \ref{S:C}. 

\section{Genuine multipartite entanglement}\label{S:GME}

In this section we briefly review the concept of genuine entanglement. We consider three qubits as an example to explain the concept. A state is 
separable with respect to some bipartition, say, $A|BC$, if it can be written as
\begin{equation}
\rho = \sum_k \, q_k \, |\phi_A^k \rangle\langle \phi_A^k| \otimes |\psi_{BC}^k \rangle\langle \phi_{BC}^k| \, ,  
\end{equation}
where $q_k$ form a probability distribution. We denote these states as $\rho_{A|BC}^{sep}$. The two other possibilities are 
$\rho_{B|AC}^{sep}$ and $\rho_{C|AB}^{sep}$. Then a state is called biseparable if it can be written as a mixture of states which are separable 
with respect to different bipartitions, that is 
\begin{eqnarray}
\rho^{bs} = p_1 \, \rho_{A|BC}^{sep} + p_2 \, \rho_{B|AC}^{sep} + p_3 \, \rho_{C|AB}^{sep}\,.
\label{Eq:BS}
\end{eqnarray}
Any state which is not biseparable is called genuinely entangled. The description of genuine entanglement is quite challenging. Considerable efforts have been 
devoted for its characterization, quantification, detection and preparation 
\cite{Sabin-EPJD48-2008, A-Kay, Guehne-NJP12-2010, Guehne-PLA375-2011, Duer-PRA61-2000, Acin-PRL87-2001, Monz-PRL106-2011}.

Biseparable states can be created by entangling any two of three particles and then one can create a statistical mixture by forgetting to which pairs this 
operation was applied. To detect genuine entanglement, it is not enough to apply bipartite criterion to every partition, instead one has to show that it can not 
be written in form of Eq.(\ref{Eq:BS}). As there is no efficient way to search through all possible decompositions, we can consider a superset of the set of separable 
states which can be characterized more easily than the set of separable states. As a superset of states that are separable with respect to, say, 
partition $A|BC$, we select the set of states that have a positive partial transpose $PPT$ with respect to partition $A|BC$. 
A state $\rho = \sum_{ijkl} \, \rho_{ij,kl} \, |i\rangle\langle j| \otimes |k\rangle\langle l|$ is PPT if its partially transposed matrix 
$\rho^{T_A} = \sum_{ijkl} \, \rho_{ji,kl} \, |i\rangle\langle j| \otimes |k\rangle\langle l|$ has no negative eigenvalues. The convex set 
$\rho_{A|BC}^{sep}$ is contained in a larger convex set of states which has a positive partial transpose. The benefit of doing so is the easy 
characterization of PPT set. A well known fact is that separable states are always PPT \cite{HHH-PLA96}.
We denote the states which are PPT with respect to fixed bipartition by $\rho_{A|BC}^{ppt}$, $\rho_{B|AC}^{ppt}$, and $\rho_{C|AB}^{ppt}$, and 
ask the question that whether a state can be written as
\begin{eqnarray}
\rho^{pptmix} = p_1 \, \rho_{A|BC}^{ppt} + p_2 \, \rho_{B|AC}^{ppt} + p_3 \, \rho_{C|AB}^{ppt}\,.
\end{eqnarray}
Such mixing of PPT states is called PPT-mixture and it can be characterized more easily than biseparable states. This method allows use of semidefinite 
programming (SDP) and also gives an entanglement montone. 
For bipartite systems, this monotone is equivalent to an entanglement measure called {\it negativity} \cite{Vidal-PRA65-2002}. For multipartite systems, 
this monotone may be called genuine negativity. For multiqubits the value of genuine negativity can be at most $1/2$ \cite{Bastian-PRL106-2011}. 

\section{Generating entanglement via XY-interactions among phase qubits}\label{S:GXY}

There are several physical systems such as spins, atoms, or photons, that might be used as quantum information processing devices. 
Another promising candidate has emerged in recent years as superconducting qubit. Instead of relying on fundamental quantum systems, these devices 
are engineered circuits that consist of many constituent atoms exhibiting collective quantum behavior. The two key features are superconductivity, 
which is a collective quantum behavior of many electrons that allows the entire circuit to be treated quantum mechanically, and the Josephson effect, 
which gives the strong non linearity required to make an effective two-level system or qubit \cite{Neeley-PhDThesis}. As we study phase qubits in 
this paper and the circuit diagram for phase qubit is shown in Figure $2.1$ of Ref. \cite{Neeley-PhDThesis}. We have realized that our studies to 
generate entanglement are mostly based on mathematical model underlying and a computable measure of genuine entanglement therefore we do not need 
any detailed descriptions of phase qubits itself. Instead of writing these details here, we refer the readers to see Ref. \cite{Neeley-PhDThesis} 
and references therein. 

\subsection{Generating entanglement for three qubits}

\begin{figure}[h]
\scalebox{1.75}{\includegraphics[width=1.6in]{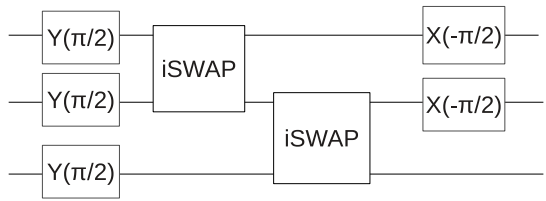}}
\centering
\caption{The circuit diagram for generating $|GHZ_3\rangle$ state using two qubit $iSWAP$ gate, which can be generated directly by capacitive 
coupling in phase qubits. Single qubit rotations are applied before and after the action of gate.}
\label{Fig:2}
\end{figure}

Let us first consider the three qubits case. An arbitrary initial pure state for three qubits can be written as 
$|\psi(0)\rangle = \sum_{ijk = 0}^1 \, C_{ijk}(0) \, |ijk\rangle$, where we have expressed the state in the computational basis 
$\{ \,|000\rangle , \, |001\rangle,  \, \ldots \,|111\rangle\}$ and $C_{ijk} (0)$ are the initial probability amplitudes. To create 
$|GHZ_3\rangle = 1/\sqrt{2} \, (|000\rangle + |111\rangle)$ state, one can design the quantum circuit diagram \cite{Neeley-Nature467-2010} 
shown in Figure \ref{Fig:2}. This Figure utilizes a more natural universal gate called {\it iSWAP} gate \cite{Schuch-PRA67-2003}. The {\it iSWAP} 
gate can be generated by applying the interaction Hamiltonian for the superconducting phase qubits written as 
\cite{Neeley-Nature467-2010, Schuch-PRA67-2003}  
\begin{eqnarray}
H_{int}^{ij} = \frac{\hbar g}{2} \left( \sigma_x^i \sigma_x^j + \sigma_y^i \sigma_y^j \right)\,, 
\label{Eq:IH}
\end{eqnarray}
for interaction time $t_{iSWAP} = \pi/(2g)$, where $g$ is the coupling strength and $\sigma_x,\, \sigma_y$ are the Pauli operators on qubits $i$ 
and $j$. Alternatively, this coupling Hamiltonian can be written as
\begin{eqnarray}
H_{int}^{ij} = \hbar \, g \, \left( \sigma_+^i \sigma_-^j + \sigma_-^i \sigma_+^j \right)\,, 
\label{Eq:IHA}
\end{eqnarray}
where $\sigma_+$ and $\sigma_-$ are the operators which create and destroy an excitation in each qubit respectively. This form reflects the fact that 
the interaction leads to excitation swapping back and forth between the two coupled qubits. Under the specified interaction time, qubit states are 
transform as $| 0 1 \rangle \mapsto -{\rm i} | 1 0 \rangle$ and $| 1 0 \rangle \mapsto -{\rm i} | 0 1\rangle $ whereas $|00\rangle $ and $|11\rangle$
invariant. It is simple to figure out that the circuit shown in Figure \ref{Fig:2} produce $|GHZ_3\rangle$ state provided all three initial qubits 
are in their ground state $|0\rangle$ \cite{Neeley-Nature467-2010}. This means that if we start with all qubits in the ground state, 
apply rotation $Y(\pi/2)$ achieved by shining a laser on all three qubits, after that we  
first turn on the interaction for time $t_{iSWAP}$ between qubits $1$ and $2$ and then we turn on the interaction between qubits $2$ and $3$ 
again for time $t_{iSWAP}$, and finally we apply the rotation $X (-\pi/2)$ again by shining laser, then we get the GHZ state. 
We note that $|GHZ_3\rangle$ is maximally entangled state measured by 
genuine negativity, that is, $E (|GHZ_3\rangle) = 1/2$.

\begin{figure}[h]
\scalebox{1.75}{\includegraphics[width=1.6in]{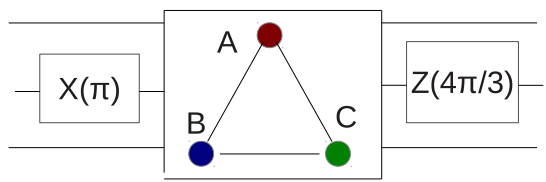}}
\centering
\caption{The circuit diagram for generating $|W_3\rangle$ state of three qubits using single entangling step with simultaneous coupling between all 
qubits. Single qubit rotations are applied on the middle qubit before and after such interaction.}
\label{Fig:3}
\end{figure}

Another inequivalent entangled state which has been created in this setup is the 
$|W_3\rangle = 1/\sqrt{3} \, ( |001 \rangle + |010\rangle + |100 \rangle)$ state \cite{Neeley-Nature467-2010}, which do not have maximum 
amount of entanglement as measured by genuine negativity, that is $E(|W_3\rangle) \approx 0.4428$. The corresponding circuit diagram for 
generating $|W_3 \rangle $ state is shown in Figure \ref{Fig:3}. It can be seen that first middle qubit is excited by applying a $\pi$-pulse to 
qubit $B$ to excite it with single photon. The effect is the creation of state $|010\rangle$. After that the interaction among all qubits can be 
turned on by applying the interaction Hamiltonian  
\begin{eqnarray}					
H_{int} = H_{int}^{AB} + H_{int}^{AC} + H_{int}^{BC} \, ,
\label{Eq:IHW}
\end{eqnarray}
for the interaction time $t_W = (4/9) \, t_{iSWAP}$ \cite{Neeley-Nature467-2010}, which corresponds to $g t_W  = (2 \, \pi)/9 \approx 0.7$. This 
action leaves all qubits in an equal superposition of single excitation. Finally a Pauli $Z$ matrix is applied to correct the phase of qubit $B$. 
The Pauli rotations $X$ and $Z$ on second qubit can be manipulated by focusing a laser before and after turning on interaction for time $t_w$.
The tomographic data \cite{Neeley-Nature467-2010} yields a very efficient state $|W_3\rangle$ as a result. In the following, we reconfirm 
this result using computable entanglement monotone $E$. In addition, we further investigate this setup and suggest that it can offer some more 
possibilities to create multipartite entangled states.    

\begin{figure}[h]
\scalebox{2.25}{\includegraphics[width=1.95in]{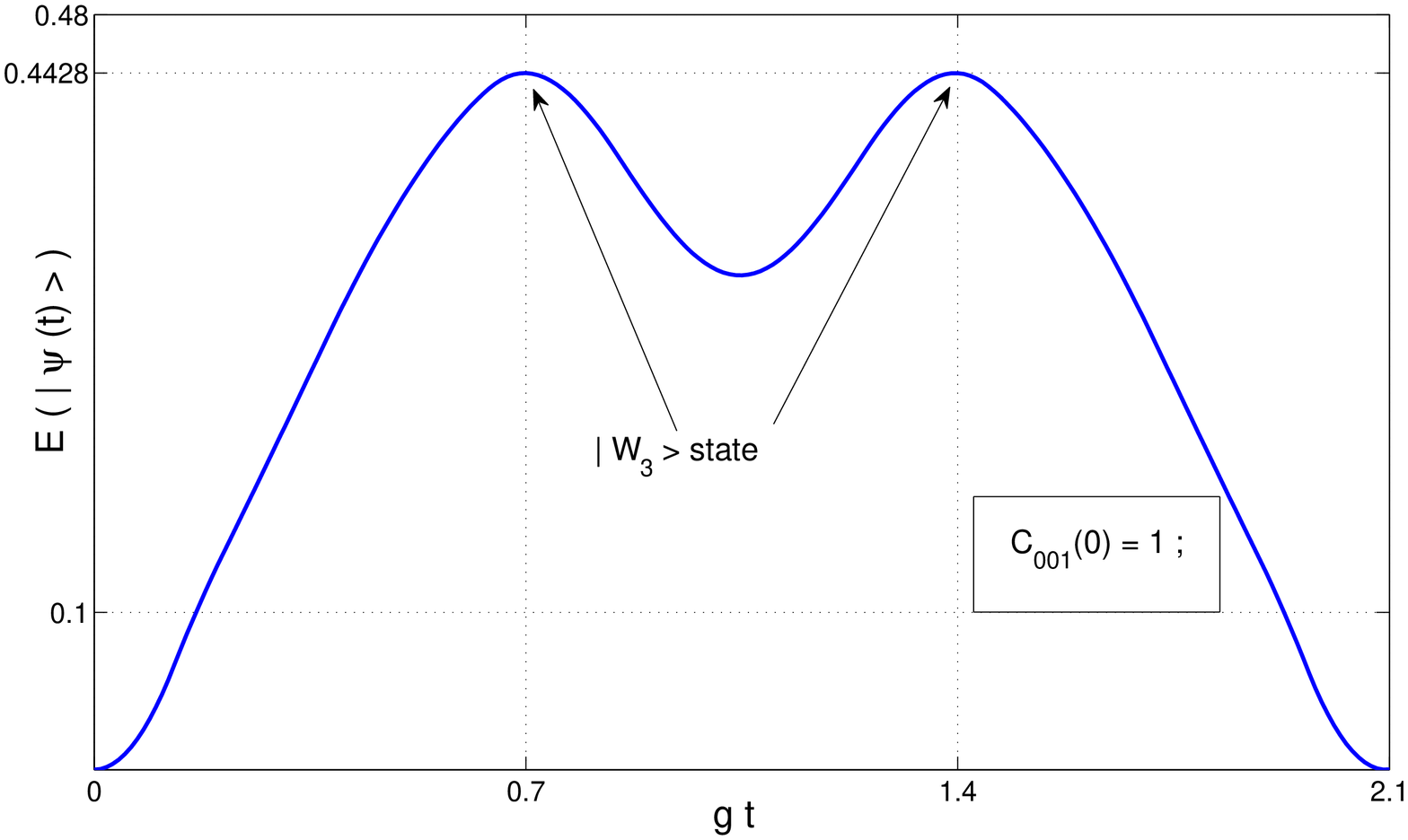}}
\centering
\caption{Generalized negativity $E(|\psi(t)\rangle)$ is plotted against the parameter $gt$ for initial condition $C_{001}(0) = 1$. It can be seen 
that $|W_3\rangle$ state is created at $gt \approx 0.7$ and $gt \approx 1.4$.}
\label{Fig:4}
\end{figure}

This Hamiltonian (\ref{Eq:IHW}) leads to two independent set of coupled equations for probability amplitudes, that is, 
$\{ \, C_{001}(t),\, C_{010}(t), \, C_{100}(t) \, \}$ and $\{ \, C_{110}(t), \, C_{101}(t), \, C_{011}(t) \, \}$, with $C_{000}(0)$ and 
$C_{111}(0)$ as invariants. The key observation is that if one has any one of nonzero amplitudes in any set then one can generate either 
$|W_3\rangle$ state or $|\widetilde{W}_3 \rangle = 1/\sqrt{3} \, ( |110 \rangle + |101\rangle + |011 \rangle)$ state. In Figure \ref{Fig:4} we 
plot $E(|\psi(t)\rangle)$ against parameter $gt$ with initial condition $C_{001}(0) = 1$ and all other $C_{ijk}(0) = 0$. We observe the creation of 
$| W_3 \rangle$ as predicted \cite{Neeley-Nature467-2010} at $gt_W \approx 0.7$ and again at $gt_W \approx 1.4$. However, to correct the phase of 
$|W_3 \rangle$ state in second peak, we need to apply $Z(8\pi/3)$ rotation.

We investigate the possibility to create the $|G_3\rangle$ states \cite{Sen-PRA68-2003} which are defined as
\begin{eqnarray} 
|G_N^\pm \rangle = (|W_N\rangle \pm |\widetilde{W}_N\rangle)/\sqrt{2}\,.
\label{Eq:GN}
\end{eqnarray}
One interesting feature of $|G_3\rangle$ states is the possibility of transforming them directly into $|GHZ_3\rangle$ state via local filters 
given by \cite{Wieczorek-PRA79-2009}
\begin{eqnarray}
f_+ =& \mathcal{H} \, \bigg\{ \frac{1}{2} \bigg[ \bigg( \frac{1}{\sqrt{3}} + {\rm i} \bigg) I + \bigg( \frac{1}{\sqrt{3}} 
- {\rm i} \bigg) \sigma_z \, \bigg] \bigg\} \mathcal{H} \,,\nonumber \\ 
f_- =& \mathcal{H} \, \bigg\{ \frac{1}{2} \bigg[ \bigg( \frac{1}{\sqrt{3}} + {\rm i} \bigg) \sigma_x + {\rm i} \, \bigg( \frac{1}{\sqrt{3}} 
- {\rm i} \bigg) \sigma_y \, \bigg] \bigg\} \mathcal{H} \,, 
\end{eqnarray}
where $\sigma_i$ are the Pauli matrices and $\mathcal{H}$ is the Hadamard transformation. As a result, we get
\begin{eqnarray}
(f_+ \otimes f_+ \otimes f_+) \,|G_3^+\rangle  = \frac{1}{3} |GHZ_3\rangle \,,\nonumber \\
(f_- \otimes f_- \otimes f_-) \,|G_3^-\rangle  = \frac{1}{3} |GHZ_3\rangle \, ,
\end{eqnarray}
with probability $1/9$. Genuine negativity for $|G_3\rangle$ is $E(|G_3\rangle) \approx 0.3448$. One possibility to create $|G_3\rangle$ seems 
straight forward from two independent sets of probability amplitudes. By having excitations in each one of this set, the experiment performed for 
$|W_3\rangle$ state \cite{Neeley-Nature467-2010} may lead to $|G_3\rangle$ state at $gt \approx 0.7$. However, this is not what we actually observe 
with initial conditions $C_{001}(0) = C_{011}(0) = 1/\sqrt{2}$. Although at $gt \approx 0.7$ both $|W_3\rangle$ and $|\widetilde{W}_3\rangle$ are 
formed, nevertheless their respective phases also mix which we were able to correct only in case of $|W_3\rangle$ state creation. Hence the state 
generated at $gt \approx 0.7$ is 
\begin{eqnarray}
|\psi (gt \approx 0.7) \rangle = \frac{1}{\sqrt{2}} \, ( {\rm e}^{{\rm i} \phi_1(\phi_2)} \, |W_3\rangle 
+ {\rm e}^{{\rm i} \phi_2(\phi_1)} |\widetilde{W}_3\rangle).  
\end{eqnarray}
Due to this phase dependence, the perfect $|G_3 \rangle$ can not be generated in this method. Another method is to create $|W_3\rangle$ and 
$|\widetilde{W}_3\rangle$ separately as shown above and then prepare the superposition.  

\subsection{Generating entanglement for four qubits}

Four qubits setup is an extension of three qubits setup which offers some unique possibilities to generate important multipartite entangled states. 
The interaction Hamiltonian for four body interactions can be written as
\begin{eqnarray}
H_{int} = H_{int}^{AB} + H_{int}^{AC} + H_{int}^{AD} + H_{int}^{BC} + H_{int}^{BD} + H_{int}^{CD}\,.  
\label{Eq:IH4Qb}
\end{eqnarray}
As a result of this Hamiltonian the following three sets of coupled differential equations for probability amplitudes emerge, that is, 
$\{ \, C_{0001}(t)$, $C_{0010}(t)$, $C_{0100}(t)$, $C_{1000}(t) \, \}$, $\{ \, C_{1110}(t)$, $C_{1101}(t)$, $C_{1011}(t)$, $C_{0111}(t)\, \}$, and 
$\{ C_{0011}(t)$, $C_{0101}(t)$, $C_{0110}(t)$, $C_{1001}(t)$, $C_{1010}(t)$, $C_{1100}(t) \, \}$. Whereas $C_{0000}(0)$ and $C_{1111}(0)$ are 
invariant. We recognize that the first two sets provide the possibility to generate $|W_4\rangle$ and $|\widetilde{W}_4\rangle$ states, 
respectively. Whereas the third set can be utilized to create the singlet state of four qubits as we demonstrate below.

We start with $|GHZ_4 \rangle = 1/\sqrt{2} \, (|0000 \rangle + |1111\rangle)$ state with genuine negativity $E(|GHZ_4\rangle) = 1/2$. 
The circuit diagram in this case is simple extension of Figure \ref{Fig:2} leading to desired state. Therefore, we do not repeat the diagram and 
arguments for the preparation of GHZ state here.

Next we consider the $|W\rangle$ state of four qubits given as $|W_4 \rangle = 1/2 \, (|0001 \rangle + |0010\rangle + |0100 \rangle + |1000\rangle)$, 
which is not a maximally entangled as measured by genuine negativity, that is $E (|W_4\rangle) \approx 0.366$. The circuit diagram 
for the $|W_4\rangle$ is a generalization of Figure \ref{Fig:3}. We observed that with initial condition $C_{0001}(0) = 1$, $|W_4\rangle$ state is 
created at $g t_W = \pi/4$. To correct the phase, we need to apply $Z(\pi)$ rotation, by turning on a laser.

Four qubits $|G_4 \rangle = 1/\sqrt{2} \, (|W_4\rangle + |\widetilde{W}_4\rangle)$ state is interestingly a maximally entangled state as measured 
by genuine negativity, that is, $E(|G_4\rangle) = 1/2$. 
\begin{eqnarray}
|G_4 \rangle =& \frac{1}{2\, \sqrt{2}} \, (| 0001 \rangle + |0010 \rangle + |0100\rangle + |1000 \rangle \nonumber \\& + |0111\rangle + |1011\rangle 
+ |1101 \rangle + |1110\rangle) \, .
\end{eqnarray}
To generate this state, we start with initial condition $C_{0001}(0) = C_{0111}(0) = 1/\sqrt{2}$ and expect $|G_4\rangle$ state to be created at 
$gt = \pi/4 $. However, after correcting phase, the state we obtain is 
\begin{eqnarray}
|\psi_G(gt = \pi/4) \rangle =& \frac{1}{2\, \sqrt{2}} \, (| 0001 \rangle + |0010 \rangle + |0100\rangle + |1000 \rangle \nonumber \\& 
+ |0111\rangle - |1011\rangle - |1101 \rangle + |1110\rangle) \, .
\end{eqnarray}
This state is also maximally entangled as measurement by genuine negativity, that is, $E(|\psi_G \rangle) = 1/2$.

\begin{figure}[h]
\scalebox{1.95}{\includegraphics[width=1.75in]{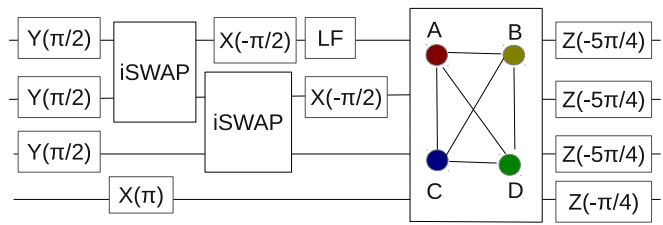}}
\centering
\caption{Circuit diagram is shown for generating $|\chi_4\rangle$ state. $|GHZ_3\rangle$ state is created in the first three qubits before 
applying a local filter and four body interactions with fourth qubit in excited state. Phases are corrected by $Z(\gamma)$ rotations at the end.}
\label{Fig:9}
\end{figure}

The question of maximally entangled states for N qubits depends on the entanglement measure being chosen \cite{Guehne-PR-2009}. 
In fact there are several inequivalent measures of multipartite entanglement, and maximally entangled states differ for different measures. 
For a class of entanglement measures, based on anti-linear operators and combs \cite{Osterloh-PRA72-2005, Osterloh-IJQI-2006}, there are 
three different measures of this type for four qubits, resulting in three different maximally entangled states \cite{Guehne-PR-2009}. 
The first one is the $|GHZ_4\rangle$ state, the second one is the cluster state, and the third one is the state
\begin{eqnarray}
|\chi_4 \rangle = \frac{\sqrt{2} \, | 1111 \rangle + |0001 \rangle + |0010\rangle + |0100 \rangle + |1000\rangle}{\sqrt{6}} \, .
\end{eqnarray}
Surprisingly, it turns out that all these states are also maximally entangled measured by genuine negativity, that is $E (|\chi_4\rangle) = 1/2$. 
It has been observed that $|\chi_4\rangle$ state is the symmetric four-qubit state that maximizes certain bipartite entanglement 
properties \cite{BKraus-2003}. In addition, it was shown \cite{Osterloh-IJQI-2006} that one can generalize this state to five and six qubits, 
where it is also a maximally entangled state for some comb measure. The generation of this state is still an experimental 
challenge \cite{Guehne-PR-2009}. We propose a simple modification of current experimental setup which seems quite feasible and 
would lead to creation of $|\chi_4\rangle$ state. To this end we realize that the state $|\chi_4\rangle$ has a large overlap with $|W_4\rangle$ 
state with additional component $|1111\rangle$. It is easy to see that $|\chi_4\rangle$ can be written as
\begin{eqnarray}
 |\chi_4\rangle = \alpha \, |1111\rangle + \sqrt{1-\alpha^2} \, |W_4\rangle \, ,
\end{eqnarray}
with $\alpha = 1/\sqrt{3}$. It is interesting to note that $|W_4\rangle$ is not maximally entangled, however its entanglement can be maximized 
by mixing the component $|1111\rangle$. One can also try to maximize entanglement of $|W_3\rangle$ by mixing the component $|111\rangle$. 
It turns out that one can do that and the resulting state $|\chi_3\rangle = (|001\rangle + |010\rangle + |100\rangle - |111\rangle)/2$ is indeed 
a maximally entangled, that is $E(|\chi_3\rangle) = 1/2$, however $|\chi_3\rangle$ is equivalent to $|GHZ_3\rangle$ 
state \cite{Svetlichny-PRD35-1987}, whereas $|\chi_4\rangle$ is inequivalent to $|GHZ_4\rangle$. We propose the circuit diagram for the experiment 
to generate $|\chi_4\rangle$ in Figure \ref{Fig:9}.

\begin{figure}[h]
\scalebox{2.25}{\includegraphics[width=1.95in]{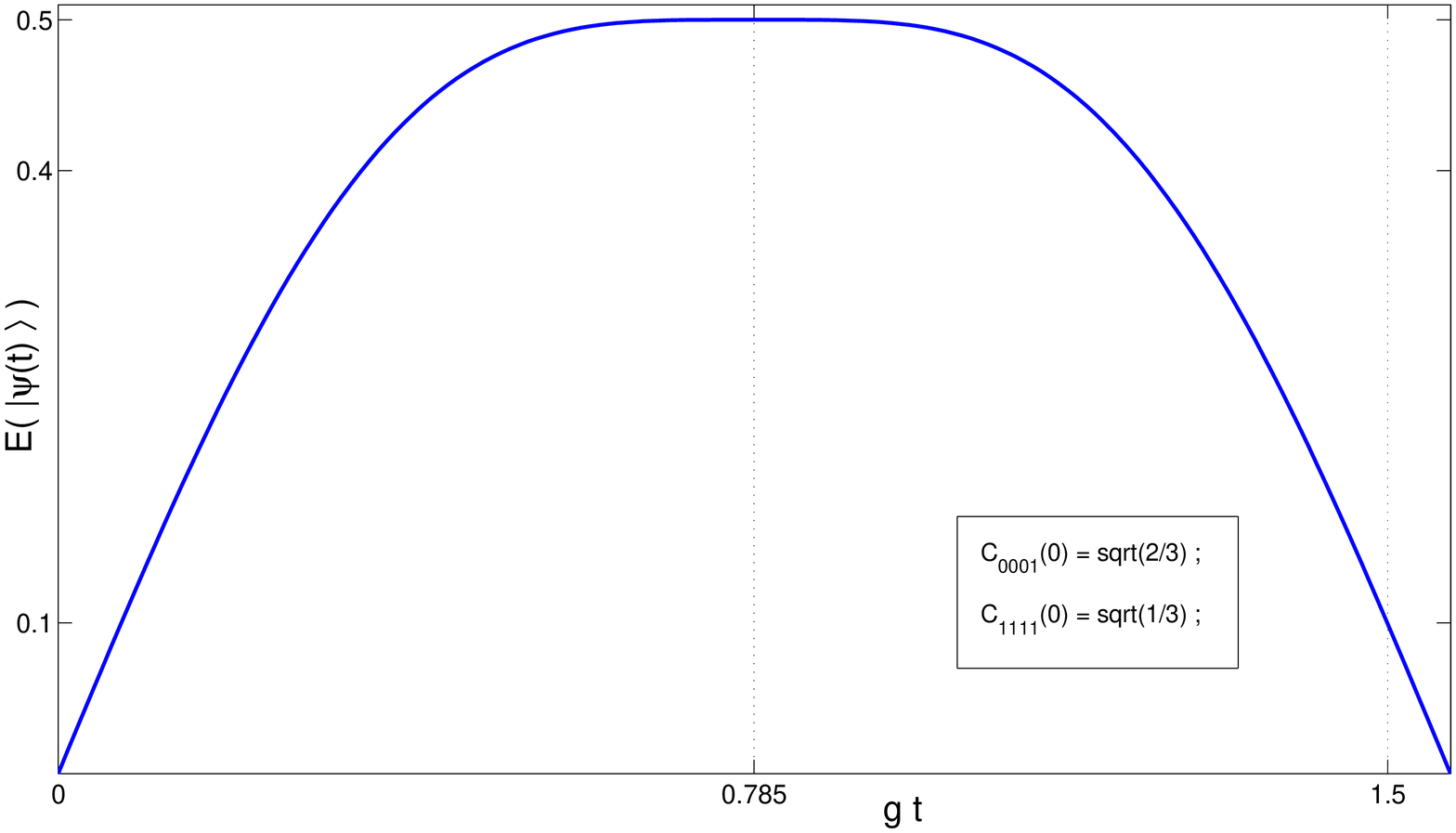}}
\centering
\caption{$E(|\psi(t)\rangle)$ is plotted against parameter $gt$ with initial conditions $C_{0001}(0) = \sqrt{2/3}$ and $C_{1111}(0) = 1/\sqrt{3}$. }
\label{Fig:10}
\end{figure}

The idea is to first generate $|GHZ_3\rangle$ state in the first three qubits using the method described in Figure \ref{Fig:2}. 
Once $|GHZ_3\rangle$ state is created then one can apply local filter (LF) on the first qubit to transform $|GHZ_3\rangle$ state into 
non-maximally entangled state 
\begin{eqnarray}
|\psi_{ABC}\rangle = \sqrt{2/3} \, |000\rangle + \sqrt{1/3} \, |111\rangle \, .
\end{eqnarray}
The precise form of the filter $f$ which does this job, is given as $f = diag \{ a, \, 1/a \} $ with $a = (2)^{1/4}$. After that an excitation is 
created on the fourth qubit and then finally this input state is allowed to interact as four body interaction for same amount of time which is 
required to create $|W_4\rangle$ state, that is, $g t_W = \pi/4 $. It is not difficult to see that this procedure is equivalent to 
produce the initial conditions on probability amplitudes such that $C_{0001} (0) = \sqrt{2/3}$ and $C_{1111}(0) = 1/\sqrt{3}$. 
In Figure \ref{Fig:10}, we have plotted $E(|\psi(t) \rangle)$ against parameter $gt$ with these initial conditions. It can be seen that 
$|\chi_4\rangle$ state is created as genuine negativity achieves its maximum value at $gt = \pi/4 \approx 0.785$.
To correct phases, we first need to apply $Z(\pi)$ rotation on the fourth qubit as we did in $| W_4 \rangle$ case. As $|1111\rangle$ component is 
in the superposition, so this operation introduces some additional phases and to correct them we need to apply $Z(-5 \pi/4)$ rotations on all four 
qubits. The resultant rotation on the fourth qubit is $Z(\pi \, - \, 5 \pi/4) \, = \, Z(-\pi/4)$.

\begin{figure}[h]
\scalebox{2.15}{\includegraphics[width=1.95in]{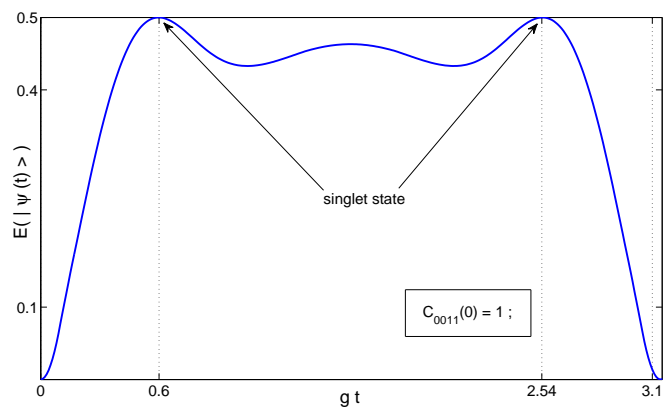}}
\centering
\caption{$E(|\psi(t)\rangle)$ is plotted against parameter $gt$ with initial conditions $C_{0011}(0) = 1$. The singlet state is generated at 
$g t \approx 0.6$ and at $g t \approx 2.54$. }
\label{Fig:11}
\end{figure}

Our final example is the possibility of creating the multiqubit singlet state. These are pure states $|\Psi\rangle$ which are invariant under a 
simultaneous unitaries on all qubits. Such states only exist for an even number of qubits. For two qubits, singlet state is given as 
$|\psi^- \rangle = 1/\sqrt{2} (|01\rangle - |10\rangle)$. For four qubits, singlet state \cite{Guehne-PR-2009, Weinfurter-PRA64-2001} is given as 
\begin{eqnarray}
|\Psi_{S,4}\rangle =& \frac{1}{\sqrt{3}} \, \big[ \, |0011 \rangle + |1100\rangle - 1/2 \, ( \, |0101 \rangle \nonumber\\& + |0110\rangle  
+ |1001 \rangle + |1010\rangle \, ) \, \big]\, , 
\end{eqnarray}
with $E (|\Psi_{S,4}\rangle) = 1/2$, which means that this state is also maximally entangled state according to genuine negativity. The 
possibility to create this state appears natural as the third set contains all relevant probability amplitudes for singlet state. 
One could work out the corresponding circuit diagram for this case but the essential point is the fact that we need 
at least one non-zero probability amplitude only in the third set. It follows that the resulting dynamics leads to the singlet state at specific 
times. Figure \ref{Fig:11} shows the evolution of $E(|\psi(t)\rangle)$ against parameter $gt$ with initial condition $C_{0011}(0) = 1$. It can be 
seen that $|\Psi_{S,4}\rangle$ is created at $g t \approx 0.6$ and at $gt \approx 2.54$. We emphasize at this point that although we have plotted 
the numerical value of $E(|\psi(t)\rangle)$ for all the cases studied in this paper, however the detailed analysis of probability amplitudes 
confirm that the corresponding states are created at specific instances of time with specific initial conditions. 

\section{Generating multipartite entanglement via von Neumann projections}\label{S:GVN}

In this section, we review the idea of generating multipartite entanglement via applying von Neumann projective measurements on some of the $n$-qubit 
state to get the $m$-qubit state with $n > m$. We stress here at this point that this idea is not new and extensive work has been done in 
linear optical quantum computation \cite{Kiesel-PRL98-2007, Wieczorek-PRA79-2009, Wieczorek-PRL103-2009}. Our main purpose here in this paper is to 
compare this method with previous section in order to show the interesting correspondence between inequivalent genuine multipartite entangled 
states. Let $\rho_n$ be an initial $n$-qubit density matrix and $\rho_m$ be the density matrix of $m$ qubits after applying projective measurements 
on the $n-m$ qubits. Mathematically, one can write this operation as
\begin{eqnarray}
\fl \rho_m = \frac{1}{\mathcal{N}} \Tr_{ij} \, [ \,( I \otimes \ldots M_i \otimes \ldots M_j \otimes \ldots I) \, \rho_n \, (I \otimes \nonumber \\ 
\ldots M_i^\dagger \otimes \ldots M_j^\dagger \otimes \ldots  I) \, ] \,,
\end{eqnarray}
where $\mathcal{N}$ is the normalization factor, $\Tr_{ij}$ is the partial trace over qubits being measured, and $M_i = V \Pi_i V^\dagger$ is the 
von Neumann projection operator on $i$th qubit, with 
$\Pi_i = | i \rangle\langle i|$ is the projector and $V = t \, I + i \, \vec{y}\, \cdot \, \vec{\sigma}$ is the unitary matrix, that is, 
$V \in SU(2)$, such that $t, y_i \in \mathcal{R}$ and $t^2 + y_1^2 + y_2^2 + y_3^2 = 1$ 
and $\sigma$ is a vector of Pauli matrices.

In recent experimental work \cite{Kiesel-PRL98-2007, Wieczorek-PRA79-2009, Wieczorek-PRL103-2009}, the authors mainly considered this 
problem for Dicke states of four, five and six qubits and experimentally demonstrated the possibility of obtaining inequivalent multipartite 
entangled states by projective measurements. In this work, we examine the several inequivalent multipartite entangled states for four qubits and 
investigate the possibility to obtain\footnote{Here the word obtain means that we get the result after the arrow by applying projective measurements and tracing out 
the qubit being measured.} inequivalent multipartite entangled states for three qubits. We have chosen only those examples which are not considered before. 
%

Our first example is the cluster state which is one of the two independent graph states for four qubits. The other independent graph state for 
four qubits is $|GHZ_4\rangle$ state \cite{Guehne-PR-2009}. The cluster state can be written in the computational basis as  
\begin{eqnarray}
 |CL_4\rangle  = \frac{1}{2} (|0000\rangle + |0011\rangle + |1100\rangle - |1111\rangle)\,.
\end{eqnarray}
Interestingly, it turns out that the cluster state can only be mapped to either one of the $|GHZ\rangle$ state of three qubits, that is  
\begin{eqnarray}
|CL_4 \rangle  \longmapsto \, \sum_{k=1}^4 \, \alpha_k \, |GHZ^k\rangle_3 \, ,
\end{eqnarray}
where any other three qubit $GHZ$ state is locally equivalent to $|GHZ_3\rangle$ state. It is known that $|GHZ_3\rangle$ is the only independent 
graph state for three qubits \cite{Guehne-PR-2009}.


The next example is the four qubit singlet state $|\Psi_{S, 4}\rangle$ which can also be mapped to superposition of $|W_3\rangle$ and 
$|\widetilde{W}_3\rangle$ state, that is 
\begin{eqnarray}
|\Psi_{S,4}\rangle \, \longmapsto \, \alpha_6 \,|W_3 \rangle + \beta_6 \, |\widetilde{W}_3\rangle \, , 
\end{eqnarray}
which can further be converted into $|G_3\rangle$ state and subsequently into $|GHZ_3\rangle$ state as discussed before.

Our final example is the $|\chi_4\rangle$ state which is the most interesting and illustrative case. We saw in the previous section that 
we combined $GHZ$-type entanglement and $W$-type entanglement to create this state. In the reverse process of projective measurements, we find 
that $|\chi_4\rangle$ state can be mapped to superposition of these two types of states, that is 
\begin{eqnarray}
|\chi_4 \rangle \, \longmapsto \, \alpha_7 \,|W_3 \rangle + \beta_7 \, |\widetilde{GHZ}_3\rangle\,, 
\end{eqnarray}
where one can transform the non maximally entangled $|\widetilde{GHZ}_3\rangle = \sqrt{1/3} \, |000\rangle + \sqrt{2/3} \, |111\rangle $ into 
$|GHZ_3\rangle$ by local filtering. All these examples have increased our understanding about the structure of genuine multipartite entangled 
states. We have seen that $|\chi_4\rangle$, $|GHZ_4\rangle$, and $|W_4\rangle$ states are all inequivalent states, however, it is interesting to 
find that one can create another inequivalent entangled states by combining two different inequivalent states.

\section{Conclusions}\label{S:C}

We have studied the generation of multipartite entangled states for superconducting phase qubits. We have shown that
this model has the capacity to generate many important multipartite entangled states. For three qubits, we have reconfirmed the previous studies on 
creating $|W_3\rangle$ state and have examined the possibility to generate $|G_3\rangle$ states.  
We have examined the possibility to create several entangled states for four qubits. In fact, the experimental setup has already been designed for 
four qubits with studies only on three qubits \cite{Neeley-Nature467-2010}. All cases studied in this paper may be performed in exactly the same 
experimental setup designed for generating $|W_3\rangle$ state \cite{Neeley-Nature467-2010}. We have shown that four qubit setup is richer than 
three qubits and offer some unique possibilities to create several important entangled states. Particularly, we have proposed the experimental 
architecture for generating $|\chi_4\rangle$ state, which seems quite feasible in this setup. Indeed the experimental creation of $|\chi_4\rangle$ 
state is challenging and such verification would be of much interest and importance for applications of this state in quantum information. We have 
also reexamined an alternative approach to generate genuine multipartite entangled states via von Neumann projective measurements. In this technique, 
one can apply projective measurements on one or more qubits of a higher dimensional density matrix before tracing them out to obtain multipartite 
entangled states in a lower dimensional space. We have only restricted ourselves to four qubits by applying projective measurements only on a single 
qubit. We have shown that we can generate both $GHZ$-type and $W$-type entangled states on a three qubit space. We have also shown that 
$|\chi_4\rangle$ state can be mapped to superposition of $GHZ$-type and $W$-type entanglement which is intuitive as we actually utilized both these 
types of entanglement to generate it. Hence this study has increased our understanding of the structure of genuine multipartite entangled states. 
One of the future avenues is to investigate the robustness of this setup against decoherence.

\ack
The author would like to thank Andreas Osterloh for discussions, Otfried G\"uhne for his generous hospitality at the 
University of Siegen, and Geza T\'oth for his kind hospitality at the University of Basque Country, Bilbao, where this work 
was orally presented. This work has been supported by the EU (Marie Curie CIG 293993/ENFOQI) and the BMBF (Chist-Era Project QUASAR).
The author is also grateful to referees for their positive comments which improved the earlier manuscript.

\end{document}